# Dynamical Models of Stock Prices Based on Technical Trading Rules
# Part I: The Models

Li-Xin Wang

*Abstract*—In this paper we use fuzzy systems theory to convert the technical trading rules commonly used by stock practitioners into excess demand functions which are then used to drive the price dynamics. The technical trading rules are recorded in natural languages where fuzzy words and vague expressions abound. In Part I of this paper, we will show the details of how to transform the technical trading heuristics into nonlinear dynamic equations. First, we define fuzzy sets to represent the fuzzy terms in the technical trading rules; second, we translate each technical trading heuristic into a group of fuzzy IF-THEN rules; third, we combine the fuzzy IF-THEN rules in a group into a fuzzy system; and finally, the linear combination of these fuzzy systems is used as the excess demand function in the price dynamic equation. We transform a wide variety of technical trading rules into fuzzy systems, including moving average rules, support and resistance rules, trend line rules, big buyer, big seller and manipulator rules, band and stop rules, and volume and relative strength rules. Simulation results show that the price dynamics driven by these technical trading rules are complex and chaotic, and some common phenomena in real stock prices such as jumps, trending and self-fulfilling appear naturally.

## I. Introduction

Understanding the dynamics of stock prices is one of the hardest challenges to human intelligence[1]. There are mainly three approaches to studying the *dynamic* changes of stock prices [2] : the random walk model (Bachelier, 1900), the agent-based models (Tesfatsion and Judd, 2006), and technical analysis (Kirkpatrick and Dahlquist, 2011).

The random walk model is the benchmark and starting point of the mainstream academic research on stock dynamics in finance and economy. At least two Nobel Prizes were awarded to the researches that used the random walk model as the starting point: the 1997 Nobel Prize of Economics to the option pricing model of Black and Scholes (Black and Scholes, 1973), and the 2003 Nobel Prize of Economics to the ARCH model of Engle (Engle, 1982). The basic assumption of the random walk model is that price changes (returns) are random and independent, so that past prices are useless in predicting the future price changes. The advantage of the random walk model is that it is a good first-approximation to real stock markets and provides a simple model based on which other important problems in finance and economy, such as option pricing (Hull, 2009) and volatility modeling (Engle, 1982; Bollerslev, Chou and Kroner, 1992; Fouque, Papanicolaou and Sircar, 2000; Andersen, Bollerslev and Diebold, 2010), can be studied in a mathematically rigorous fashion. A problem of the random walk model is that it does not provide a framework to study the microstructures of the price formation mechanism.

The agent-based models overcome this problem by following a bottom-up approach to model directly the operations of different types of traders such as fundamentalists and chartists (Hommes, 2006). There is a large literature of agent-based models for stock prices (Tesfatsion and Judd, 2006), ranging from simple heuristic models (e.g., Kyle, 1985) to sophisticated switching among different types of traders (e.g., Lux, 1998). The advantage of the agent-based computational approach is that different types of traders and various trading details can be incorporated into the models and extensive simulations can be performed to study the resulting price dynamics. The main problem of the existing agent-based models is that there are too many parameters and too many degrees of freedom such that it is difficult to determine the causes for the price features observed. This complexity also makes it very hard to calibrate the models to real stock prices. As concluded in Hommes 2006: "Simple and parsimonious

---

Li-Xin Wang is with the Department of Automation Science and Technology, Xian Jiaotong University, Xian, P.R. China (e-mail: lxwang@mail.xjtu.edu.cn).

[1] Quoting Newton's famous saying: "I can predict the motions of the heavenly bodies, but not the madness of people." (Wikiquote: Isaac Newton.)

[2] Fundamental analysis (Graham and Dodd, 1940) studies the static or equilibrium values of stock prices and therefore will not be discussed here.

heterogeneous agent models can thus help to discipline the wilderness of agent-based modeling."

The third approach to studying the dynamic changes of stock prices is technical analysis that summarizes the trading heuristics from stock practitioners over hundreds of years of real stock trading experiences in human civilization (Lo and Hasanhodzic, 2010). The foundation of technical analysis is the belief that prices follow trends so that past price information is useful in predicting the future price values. The mainstream academics do not view technical analysis as a serious scientific discipline and claim that "chart-reading must share a pedestal with alchemy" (Malkiel, 2012). While the academics stand firm on the Efficient Markets Hypothesis (Fama, 1970) and believe that stock prices are unpredictable, the practitioners are using technical trading rules to "ride the trends" to make a lot of money (Lo and Hasanhodzic, 2009). Some open-minded academics studied the performance of some technical trading rules and gave positive conclusions (e.g., Brock, Lakonishok and LeBaron, 1992; Gencay, 1998; Sullivan, Timmermann and White, 1999; Lo, Mamaysky and Wang, 2000). A main criticism of technical analysis is that most technical rules require subjective judgment because the rules are expressed in terms of natural languages where fuzzy words and vague descriptions are everywhere. For example, one such trading rule might be: "If the momentum of the uptrend is too strong, over-bought may have occurred so that a trailing stop should be placed somewhere below the current price to secure some profits." How strong is "too strong"? Where is "somewhere"? What amount should be placed at the trailing stop? This is where the academics leave the practitioners (Gigerenzer, 1996; Gigerenzer and Gaissmaier, 2011), and this is where fuzzy systems theory (Zadeh, 1971) comes to play, as we will do in this paper.

The basic idea of this paper is to use fuzzy systems theory to convert the technical trading rules into excess demand (demand minus supply) functions which are then used to drive the price dynamics. Specifically, consider a stock whose price at time point $t$ is $p_t$, where t=0,1,2,…. Suppose there are $M$ groups of traders who are trading this stock from $t$ to $t+1$, and the traders in a group are using the same technical trading rules. By converting the technical trading rules for a group, say group $i$, into a single excess demand function $ed_i(x_t)$, we obtain the price dynamical model as follows:

$$\ln(p_{t+1}) = \ln(p_t) + \sum_{i=1}^{M} a_i(t)\, ed_i(x_t) \qquad (1)$$

where the coefficient $a_i(t)$ accounts for the strength of the traders in group $i$ for the relative change of price from $t$ to $t+1$, and the $x_t$ denotes variables computed from the past prices and other information available at time $t$ (the $x_t$'s will be precisely defined in the following sections of this paper). The $a_i(t)$'s are time-varying coefficients (representing the evolving market reality (Lo, 2004)), and $a_i(t) = 0$ means the $i$'th excess demand $ed_i(x_t)$ is not present at time $t$. The $a_i(t)$'s can be determined based on stock price data $\{p_{t+1}, p_t, p_{t-1}, ...\}$, and we will show the details in Part III of this paper. To convert the technical trading rules into excess demand functions, we first define fuzzy sets to characterize the words used in the technical trading rules so that these technical rules become fuzzy IF-THEN rules in the framework of fuzzy systems theory. Then, standard fuzzy logic principles are employed to combine these fuzzy IF-THEN rules into fuzzy systems which are the excess demands $ed_i(x_t)$ in the price dynamic equation (1).

Part I of this paper is organized as follows. In Sections II, III, IV, VI and VII, we will convert a number of popular technical trading rules into excess demand functions, including moving average rules, support and resistance rules, trend line rules, band and stop rules, and volume and relative strength rules. To make the picture more complete, in Section V we will formulate the typical actions of the big buyers and big sellers (the institutional traders who manage large sums of money) and the manipulators in terms of fuzzy IF-THEN rules. Simulations will be performed in these sections to illustrate the typical price trajectories of these models. Some concluding remarks will be drawn in Section VIII.

In Part II of this paper, we will analyze the price dynamic model (1) with moving-average rules in details to show the stability, volatility, short-term predictability, return independency, fat-tailed return distribution and other properties of the model. In Part III of this paper, we will show how to detect big buyers and big sellers in Hong Kong stock market based on the price dynamic model (1) and develop two trading strategies (called follow-the-big-buyer and ride-the-mood) to beat the benchmark buy-and-hold strategy.

## II. MOVING AVERAGE RULES

The most commonly-used trading rules by technical traders are based on price moving averages of different lengths. The basic philosophy behind these rules is that the price may be on a trend if a shorter moving average is crossing a longer moving average. More specifically, one such trading heuristic is as follows.

**Heuristic 1:** A buy (sell) signal is generated if a shorter moving average of the price is crossing a longer moving average of the price from below (above). Usually, the larger the difference between the two moving averages, the stronger the buy (sell) signal. But, if the difference between the two moving averages is too large, the stock may be over-bought (over-sold), so a small sell (buy) order should be placed to safeguard the investment.

We now translate this heuristic into an excess demand function using fuzzy systems theory. The moving average of the stock price $p_t$ with length $n$ is defined as

$$\bar{p}_{t,n} = \frac{1}{n}\sum_{i=0}^{n-1} p_{t-i} \quad (2)$$

and

$$x_{1,t}^{(m,n)} = \ln\left(\bar{p}_{t,m}\big/\bar{p}_{t,n}\right) \quad (3)$$

is the log-ratio (relative change) of the price moving average of length-*m* to the price moving average of length-*n* where *m<n*. The common choices of (*m,n*) are *(1,5), (1,10), (5,20), (5,50)*, etc. (when *m=1*, $\bar{p}_{t,m}$ is the current price $p_t$). A positive $x_{1,t}^{(m,n)}$ implies a rising mode of the stock, whereas a negative $x_{1,t}^{(m,n)}$ signals the weakness of the stock price.

To translate Heuristic 1 into the language of fuzzy systems theory, we first define fuzzy sets to clarify the meaning of the linguistic descriptions for the price percentage change $x_{1,t}^{(m,n)}$ such as "$x_{1,t}^{(m,n)}$ is positive and large", "$x_{1,t}^{(m,n)}$ is negative and small", etc. Specifically, on the universe of $x_{1,t}^{(m,n)}$ (all values $x_{1,t}^{(m,n)}$ can possibly take, which we assume to be $(-\infty,\infty)$), define seven fuzzy sets: "Positive Small (PS)", "Positive Medium (PM)", "Positive Large (PL)", "Negative Small (NS)", "Negative Medium (NM)", "Negative Large (NL)", and "Around Zero (AZ)", whose membership functions are given in Fig. 1. For example, the membership function of fuzzy set PS (Positive Small) is

$$\mu_{PS}\left(x_{1,t}^{(m,n)}\right) = \begin{cases} 1 - |x_{1,t}^{(m,n)} - w|/w & if\ x_{1,t}^{(m,n)} \in [0, 2w] \\ 0 & otherewise \end{cases} \quad (4)$$

and for PL (Positive Large), the membership function is

$$\mu_{PL}\left(x_{1,t}^{(m,n)}\right) = \begin{cases} 0 & if\ x_{1,t}^{(m,n)} < 2w \\ (x_{1,t}^{(m,n)} - 2w)/w & if\ x_{1,t}^{(m,n)} \in [2w, 3w] \\ 1 & if\ x_{1,t}^{(m,n)} > 3w \end{cases} \quad (5)$$

where *w* is a positive constant. The value of *w* determines how small is small, how medium is medium, and how large is large when we use these words to describe the price changes. For example, *w =0.01* means that a change somewhere around 1% from $\bar{p}_{t,n}$ to $\bar{p}_{t,m}$ (n>m) means "small" in the trader's mind, and, "medium" means around 2%, "large" means around and above 3%. We will see later that *w* is an important parameter to shape the dynamics of the price changes.

Next, we define fuzzy sets to specify the meaning of the linguistic descriptions about the buy (sell) signals. Let *ed* be the strength of the buy or sell signal, where *ed* stands for excess demand (demand minus supply) and it can be positive (demand > supply) or negative (demand < supply). Define seven fuzzy sets "Buy Small (BS)", "Buy Medium (BM)", "Buy Big (BB)", "Sell Small (SS)", "Sell Medium (SM)", "Sell Big (SB)" and "Neutral (N)" for *ed* with membership functions shown in Fig. 2. For example, the membership function of BM (Buy Medium) is

$$\mu_{BM}(ed) = \begin{cases} (ed - 0.1)/0.1 & if\ ed \in [0.1, 0.2] \\ (0.4 - ed)/0.2 & if\ ed \in [0.2, 0.4] \\ 0 & otherwise \end{cases} \quad (6)$$

The numbers 0.1, 0.2 and 0.4 in Fig. 2 indicate, respectively, 10%, 20% and 40% of the buying or selling power of the traders who use Heuristic 1 as their trading strategy. So "Buy Small", "Buy Medium" and "Buy Big" means using around 10%, 20% and 40% of the buying power, respectively. The sell-side is just the mirror of the buy-side. You can of course choose other numbers to reflect the preference of different traders.

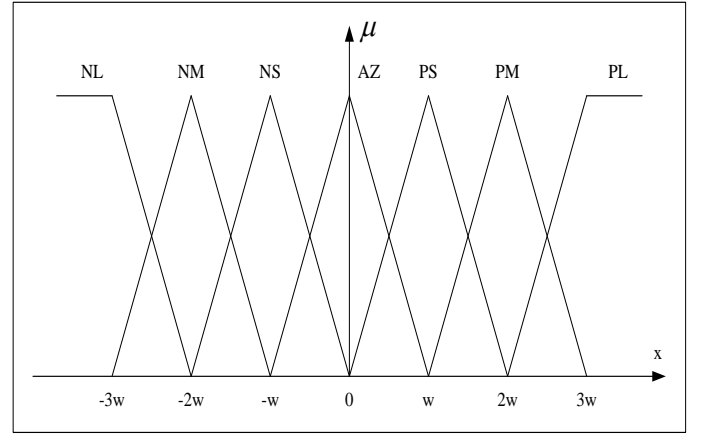

Fig. 1: Membership functions of fuzzy sets Positive Small (PS), Positive Medium (PM), Positive Large (PL), Negative Small (NS), Negative Medium (NM), Negative Large (NL) and Around Zero (AZ).

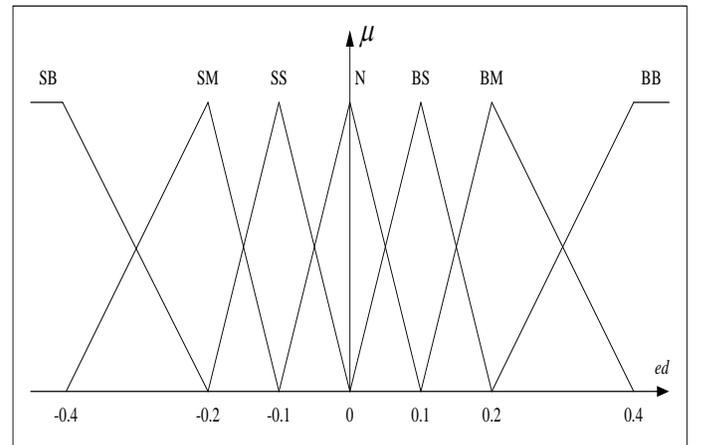

Fig. 2: Membership functions of fuzzy sets Buy Small (BS), Buy Medium (BM), Buy Big (BB), Sell Small (SS), Sell Medium (SM), Sell Big (SB) and Neutral (N) for excess demand *ed*.

With the fuzzy sets defined in Figs. 1 and 2, we can now convert Heuristic 1 into the following seven fuzzy IF-THEN rules which we call *Rule 1 Group*:

Rule $1_1$: IF $x_{1,t}^{(m,n)}$ is Positive Small (PS), THEN $ed_1$ is Buy Small (BS)

Rule $1_2$: IF $x_{1,t}^{(m,n)}$ is Positive Medium (PM), THEN $ed_1$ is Buy Big (BB)

Rule $1_3$: IF $x_{1,t}^{(m,n)}$ is Positive Large (PL), THEN $ed_1$ is Sell Medium (SM)

Rule $1_4$: IF $x_{1,t}^{(m,n)}$ is Negative Small (NS), THEN $ed_1$ is Sell Small (SS)  (7)

Rule $1_5$: IF $x_{1,t}^{(m,n)}$ is Negative Medium (NM), THEN $ed_1$ is Sell Big (SB)

Rule $1_6$: IF $x_{1,t}^{(m,n)}$ is Negative Large (NL), THEN $ed_1$ is Buy Medium (BM)

Rule $1_7$: IF $x_{1,t}^{(m,n)}$ is Around Zero (AZ), THEN $ed_1$ is Neutral (N)

The meaning of these rules is explained as follows: Rule $1_1$ and Rule $1_2$ (Rule $1_4$ and Rule $1_5$) follow the uptrend (downtrend), Rule $1_3$ (Rule $1_5$) exits (enters) the stock when it is over-bought (over-sold), and Rule $1_7$ takes no action if the price is oscillating horizontally. Clearly, these rules do what Heuristic 1 suggests.

With Heuristic 1 being transformed into the fuzzy IF-THEN rules in Rule 1 Group, our final task is to convert these rules into the excess demand function. Using fuzzy logic principles (Zadeh, 1973) and fuzzy systems theory (Zadeh, 1971; Wang, 1997), we combine the seven fuzzy IF-THEN rules in Rule 1 Group (7) into the following standard fuzzy system:

$$ed_1\left(x_{1,t}^{(m,n)}\right) = \frac{\sum_{i=1}^{7} c_i \mu_{A_i}\left(x_{1,t}^{(m,n)}\right)}{\sum_{i=1}^{7} \mu_{A_i}\left(x_{1,t}^{(m,n)}\right)} \quad (8)$$

where $A_1 = PS$, $A_2 = PM$, $A_3 = PL$, $A_4 = NS$, $A_5 = NM$, $A_6 = NL$, $A_7 = AZ$ are the fuzzy sets shown in Fig. 1, and $c_1 = 0.1$, $c_2 = 0.4$, $c_3 = -0.2$, $c_4 = -0.1$, $c_5 = -0.4$, $c_6 = 0.2$, $c_7 = 0$ are the centers of the fuzzy sets BS, BB, SM, SS, SB, BM and N shown in Fig. 2, respectively. Substituting (8) into (1), we obtain the price dynamic equation:

$$\ln(p_{t+1}) = \ln(p_t) + a_1(t)\, ed_1\left(x_{1,t}^{(m,n)}\right) \quad (9)$$

that shows how the price evolves if considering only the traders who use Heuristic 1 (Rule 1 Group) as their trading strategy.

We now simulate the price dynamic equation (9) with $(m,n)=(1,5)$, $w=0.01$ (1%) and $a_1(t) = 0.2$; three simulation runs were performed and the resulting price series $p_t$ are shown in Fig. 3, where the first 100 $p_t$ (t=1 to t=100) were generated by the random walk model:

$$\ln(p_{t+1}) = \ln(p_t) + \sigma\, \varepsilon_t \quad (10)$$

with $\varepsilon_t$ being independent Gaussian random variables of mean 0 and variance 1, $\sigma = 0.037$ and $p_0 = 10$; the remaining $p_t$ (t=101 to t=500) were generated by the price dynamical model (9). We see from Fig. 3 that although there is only one group of traders who were using the same trading rules (7), the resulting price dynamics were complex and chaotic. Comparing the random walk prices $p_1$ to $p_{100}$ with the fuzzy-rule-driven prices $p_{101}$ to $p_{500}$, we see that the big pictures look quite similar for these two fundamentally different models: one is random and the other is deterministic.

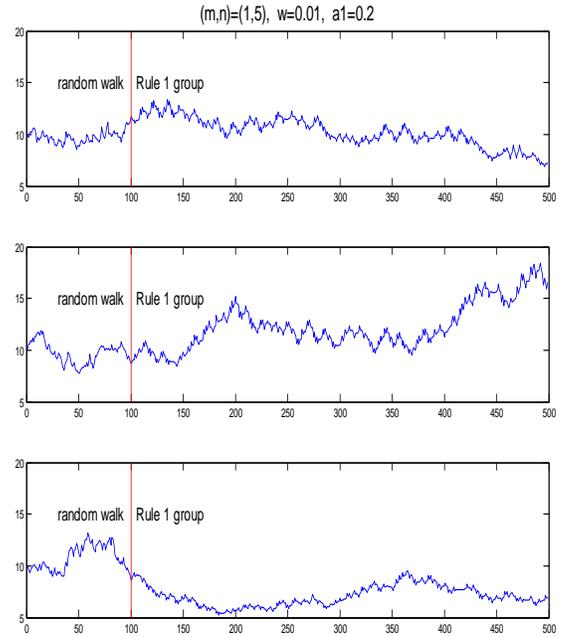

Fig. 3: Three simulation runs of price $p_t$ generated by the random walk model (10) (t=1 to t=100) and the Rule-1-Group-driven price dynamic model (9) (t=101 to t=500) with *(m,n)=(1,5), w=0.01 (1%)*, $a_1(t) = 0.03$, $\sigma = 0.037$ and $p_0 = 10$.

We now make a few remarks about what to look for from the simulated price curves (Figs. 3 to 12). Why is technical analysis so popular among traders? One explanation is that we humans have a very good vision system, and price curves provide a very convenient framework for our vision system. "A picture is worth a thousand words." The trading-heuristic-driven price dynamic models proposed in this paper produce very complex and chaotic price trajectories. In fact, whenever we made a new simulation of the price dynamic model with randomly chosen initial condition, we got a totally different price series (the three price trajectories in each Figs. 3 to 12 were chosen randomly from the simulations). This is consistent with the evolving market reality (Lo, 2004). Therefore, revealing specific characters of the models is not the main purpose of illustrating these simulated price trajectories; rather, we would like to give the reader the "big picture" or some "feeling" about what price series the models would produce. This is quite different from our usual simulations for engineering systems where some fixed phenomenon occurs whenever and wherever we do the experiment. The detailed properties of the model will be analyzed in Part II of this paper, where we will make some concrete conclusions.

### III. SUPPORT AND RESISTANCE RULES

Support and resistance are important reference points of past prices that the technical traders are looking at when they make buy or sell decisions. We first define support and resistance points. A *peak* is the price $p_k$ such that $p_k > p_{k-1}$ and $p_k > p_{k+1}$, that is, a peak is a price point that is larger than its two neighbors. A *resistance point* $resi_t^{(n)}$, defined as:

$$resi_t^{(n)} = \max_{t-n \leq k \leq t-1} \{p_k | p_k > p_{k-1}, p_k > p_{k+1}\} \quad (11)$$

is the highest peak in the time interval $[t-n, t-1]$. Similarly, a *trough* is the price $p_k$ such that $p_k < p_{k-1}$ and $p_k < p_{k+1}$, and a *support point* $supp_t^{(n)}$, defined as:

$$supp_t^{(n)} = \min_{t-n \leq k \leq t-1} \{p_k | p_k < p_{k-1}, p_k < p_{k+1}\} \quad (12)$$

is the lowest trough in the time interval $[t-n, t-1]$. When the $n$ in (11) and (12) are set equal to the current time $t$, the resistance (support) point is the highest peak (lowest trough) in the entire history of the price.

The basic philosophy of the technical trading rules based on support and resistance points is that the support (resistance) point is the lowest (highest) price in recent history where the buyers (sellers) were becoming stronger than the sellers (buyers), so that if this support (resistance) price is broken, then a downtrend (uptrend) might be establishing. So we have the following heuristic:

**Heuristic 2:** A buy (sell) signal is generated if the current price $p_t$ breaks the resistance (support) point $resi_t^{(n)}$ ($supp_t^{(n)}$) from below (above). The buy (sell) signal is weak if the breakup (breakdown) is small because a small breakup (breakdown) often ends up with a throwback (pullback). When the breakup (breakdown) becomes reasonably large, the trend is more or less confirmed so that a strong buy (sell) is recommended. If the breakup (breakdown) is too large, the stock may be over-bought (over-sold), so a small sell (buy) order should be placed to prepare for a possible reversal.

We now convert this trading heuristic into excess demand function $ed_2$ using fuzzy systems theory, similar to what we did in the last section. Define the log-ratios of the current price $p_t$ to the resistance point $resi_t^{(n)}$ and the support point $supp_t^{(n)}$ as

$$x_{2,t}^{(n)} = \ln\left(p_t / resi_t^{(n)}\right) \quad (13)$$

and

$$x_{3,t}^{(n)} = \ln\left(p_t / supp_t^{(n)}\right) \quad (14)$$

respectively. Since we are only interested in breakup ($x_{2,t}^{(n)} > 0$) and breakdown ($x_{3,t}^{(n)} < 0$) in Heuristic 2, we define fuzzy sets Positive Small (PS), Positive Medium (PM) and Positive Large (PL) as shown in Fig. 1 for $x_{2,t}^{(n)}$, and fuzzy sets Negative Small (NS), Negative Medium (NM) and Negative Large (NL), also shown in Fig. 1, for $x_{3,t}^{(n)}$. With these fuzzy sets and those defined in Fig. 2 for the buy (sell) actions, we translate Heuristics 2 into the following six fuzzy IF-THEN rules which we call *Rule 2 Group*:

Rule $2_1$: IF $x_{2,t}^{(n)}$ is Positive Small (PS), THEN $ed_2$ is Buy Small (BS)

Rule $2_2$: IF $x_{2,t}^{(n)}$ is Positive Medium (PM), THEN $ed_2$ is Buy Big (BB)

Rule $2_3$: IF $x_{2,t}^{(n)}$ is Positive Large (PL), THEN $ed_2$ is Sell Medium (SM)

Rule $2_4$: IF $x_{3,t}^{(n)}$ is Negative Small (NS), THEN $ed_2$ is Sell Small (SS)   (15)

Rule $2_5$: IF $x_{3,t}^{(n)}$ is Negative Medium (NM), THEN $ed_2$ is Sell Big (SB)

Rule $2_6$: IF $x_{3,t}^{(n)}$ is Negative Large (NL), THEN $ed_2$ is Buy Medium (BM)

Combining these rules into a standard fuzzy system we get

$$ed_2\left(x_{2,t}^{(n)}, x_{3,t}^{(n)}\right) = \frac{\sum_{i=1}^{3} c_i \mu_{A_i}\left(x_{2,t}^{(n)}\right) + \sum_{i=4}^{6} c_i \mu_{A_i}\left(x_{3,t}^{(n)}\right)}{\sum_{i=1}^{3} \mu_{A_i}\left(x_{2,t}^{(n)}\right) + \sum_{i=4}^{6} \mu_{A_i}\left(x_{3,t}^{(n)}\right)} \quad (16)$$

where $x_{2,t}^{(n)} > 0$ or $x_{3,t}^{(n)} < 0$, $A_1 = PS$, $A_2 = PM$, $A_3 = PL$, $A_4 = NS$, $A_5 = NM$, $A_6 = NL$ are the fuzzy sets shown in Fig. 1, and $c_1 = 0.1$, $c_2 = 0.4$, $c_3 = -0.2$, $c_4 = -0.1$, $c_5 = -0.4$, $c_6 = 0.2$ are the centers of the fuzzy sets BS, BB, SM, SS, SB, BM shown in Fig. 2, respectively. This $ed_2\left(x_{2,t}^{(n)}, x_{3,t}^{(n)}\right)$ is the excess demand function from all traders who use Heuristic 2 (Rule 2 Group) as their trading strategy.

Substituting $ed_1\left(x_{1,t}^{(m,n)}\right)$ of (8) and $ed_2\left(x_{2,t}^{(n)}, x_{3,t}^{(n)}\right)$ of (16) into the general model (1), we obtain the price dynamic equation

$$\ln(p_{t+1}) = \ln(p_t) + a_1(t)ed_1\left(x_{1,t}^{(m,n)}\right)$$

$$+ a_2(t)ed_2\left(x_{2,t}^{(n*)}, x_{3,t}^{(n*)}\right) \quad (17)$$

which shows how the price changes when two groups of traders are trading in the time interval *(t,t+1)*, where one group of traders use Rule 1 Group (7) and the other use Rule 2 Group (15). Fig. 4 shows the price $p_t$ for three simulation runs with parameters *(m,n)=(1,5)*, *n\*=100*, *w=0.01*, $a_1(t) = 0.2$ and $a_2(t) = 1$, where the first 100 $p_t$ (t=1 to t=100) were generated by the random walk model (10) with $\sigma = 0.005$ and $p_0 = 10$, and the remaining $p_t$ (t=101 to t=500) were generated by the price dynamic model (17).

Comparing the price trajectories in Figs. 3 and 4 we see that there were some big jumps when Rule 2 Group was added. These jumps occurred when the prices crossed the support or resistance lines.

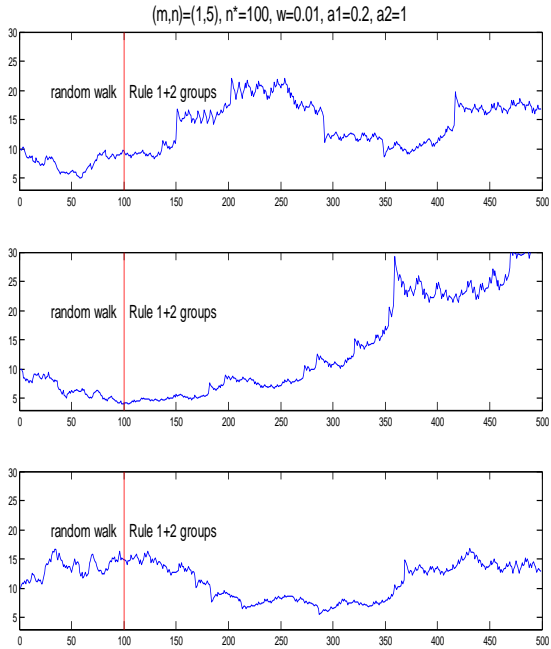

Fig. 4: Three simulation runs of price $p_t$ generated by the random walk model (10) (t=1 to t=100) and the Rule-1-plus-Rule-2-Group-driven price dynamic model (17) (t=101 to t=500) with *(m,n)=(1,5)*, *n\*=100*, *w=0.01*, $a_1(t) = 0.2$, $a_2(t) = 1$, $\sigma = 0.05$ and $p_0 = 10$.

We know that the real stock prices exhibit jumps that the random walk model cannot capture. The mathematical finance literature dealt with this problem by adding a jump random process to the random walk model (10) (e.g. Merton, 1976) or by modeling the volatility $\sigma$ as a random variable (e.g. Bollerslev, Chou and Kroner, 1992; Bouchaud and Potters, 2003; Fouque, Papanicolaou and Sircar, 2000). The problem of these random approaches is that they cannot explain why the jumps occurred at those particular locations. So it will be interesting if we can show that jumps occur when we add the excess demand $ed_2$ of Rule 2 Group to the random walk model (10), because in so doing we can explain why the jumps occurred at those particular locations: the reason is that the previous support or resistance lines were broken at these time points and traders were jumped in to catch up the trends, which caused the jumps in price. Fig. 5 shows three simulation runs of the price dynamic equation:

$$\ln(p_{t+1}) = \ln(p_t) + \sigma\,\varepsilon_t + a_2(t)ed_2\left(x_{2,t}^{(n*)}, x_{3,t}^{(n*)}\right) \quad (18)$$

with parameters *n\*=100*, *w=0.01*, $a_2(t) = 0.5$, $\sigma = 0.05$ and $p_0 = 10$, where the first 100 prices come from the pure random walk model (10) and the remaining prices are generated by (18). From Fig. 5 we see that there are indeed many jumps in the prices and these jumps are caused by the $ed_2$ in (18).

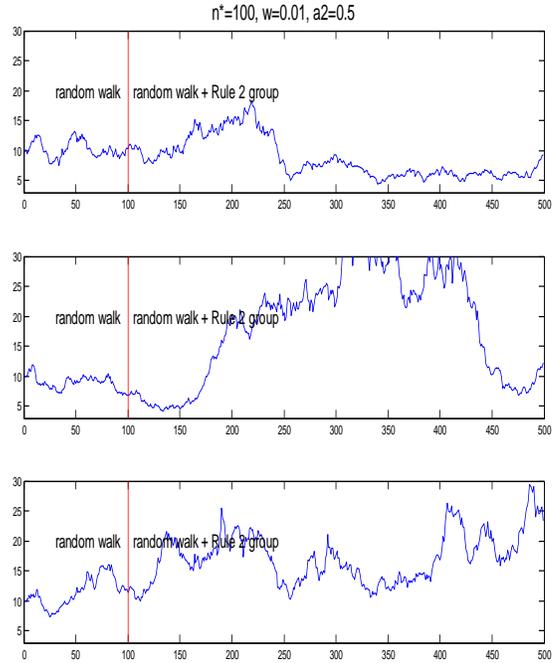

Fig. 5: Three simulation runs of price $p_t$ generated by the random walk model (10) (t=1 to t=100) and the random walk plus Rule-2-Group model (18) (t=101 to t=500) with *n\*=100*, *w=0.01*, $a_2(t) = 0.5$, $\sigma = 0.05$ and $p_0 = 10$.

Another trading heuristic related to support and resistance points concerns the traders who missed the buying (selling) opportunity at the last low (high) price. These traders have a strong tendency to buy (sell) when the price is getting around the support (resistance) point once again; this gives us the following heuristic:

**Heuristic 3:** For those traders who felt regret to miss the buying (sell) opportunity at the last low (high) price, the chance comes again when the price is getting around the support (resistance) point one more time, so buy (sell) *medium* (not *big* because the price may break across the support (resistance) line this time, and not *small* because you don't want to miss this opportunity too much).

This heuristic is transformed into the following *Rule 3 Group*:

Rule $3_1$: IF $x_{2,t}^{(n)}$ is Around Zero (AZ), THEN $ed_3$ is Sell Medium (SM)

Rule $3_2$: IF $x_{3,t}^{(n)}$ is Around Zero (AZ), THEN $ed_3$ is Buy Medium (BM)  (19)

and the excess demand function from the Rule-3-Group traders is the following fuzzy system:

$$ed_3\left(x_{2,t}^{(n)}, x_{3,t}^{(n)}\right) = \frac{0.2\, \mu_{AZ}\left(x_{3,t}^{(n)}\right) - 0.2\, \mu_{AZ}\left(x_{2,t}^{(n)}\right)}{\mu_{AZ}\left(x_{3,t}^{(n)}\right) + \mu_{AZ}\left(x_{2,t}^{(n)}\right)} \quad (20)$$

where $x_{2,t}^{(n)}$ or $x_{3,t}^{(n)} \in (-w, w)$ ($x_{2,t}^{(n)}, x_{3,t}^{(n)}$ are defined in (13), (14)), and AZ, SM and BM are shown in Figs. 1 and 2 (0.2 and -0.2 in (20) are the centers of BM and SM, respectively).

Adding all three excess demand functions $ed_1$ (8), $ed_2$ (16) and $ed_3$ (20) to the general model (1) and simulating the resulting price dynamic equation, we got Fig. 6 where the parameters were chosen as: $(m,n)=(1,5)$ in $ed_1$, the $n$'s in $ed_2$ and $ed_3$ equal $n^*=100$, $w=0.01$, $a_1 = 0.2$ and $a_2 = a_3 = 1$, and here again the first 100 $p_t$ were generated by the random walk model (10) with $\sigma = 0.05$ and $p_0 = 10$, and the remaining $p_t$ (t=101-500) were generated by the price dynamic equation (1) with $M=3$.

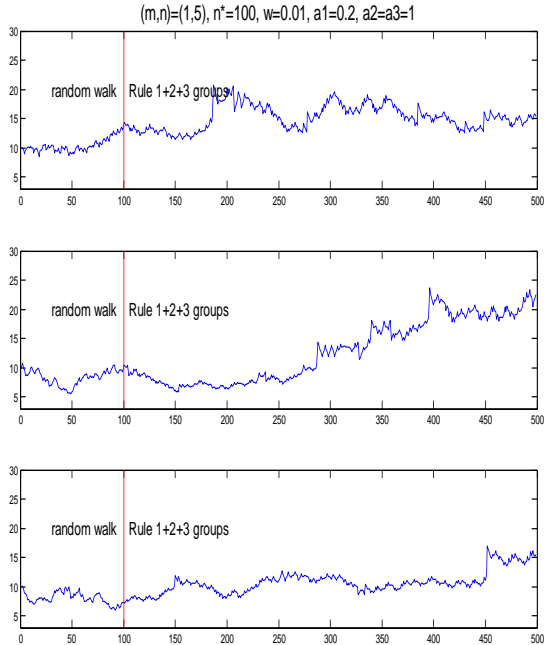

Fig. 6: Three simulation runs of price $p_t$ generated by the random walk model (10) (t=1 to t=100) and the price dynamic model (1) (t=101 to t=500) with the three excess demand functions $ed_1$, $ed_2$ and $ed_3$ from Rule 1, Rule 2 and Rule 3 Groups.

## IV. TREND LINE RULES

Trend lines are lines that connect peaks or troughs and extend into the future. There are two types of trend lines: uptrend line and downtrend line. An uptrend (downtrend) line is established if the line connecting the troughs (peaks) is pointing upwards (downwards). More specifically, let $tg_{t1}^{(n)}$ and $tg_{t2}^{(n)}$ ($pk_{t1}^{(n)}$ and $pk_{t2}^{(n)}$) be the two lowest (highest) troughs (peaks) in the time interval [*t-n, t-1*] with *t1<t2*, i.e.,

$$\left(tg_{t1}^{(n)}, tg_{t2}^{(n)}\right) = \min_{t-n \leq k \leq t-1} 2 \{p_k | p_k < p_{k-1}, p_k < p_{k+1}\} \quad (21)$$

$$\left(pk_{t1}^{(n)}, pk_{t2}^{(n)}\right) = \max_{t-n \leq k \leq t-1} 2 \{p_k | p_k > p_{k-1}, p_k > p_{k+1}\} \quad (22)$$

where *min2* (*max2*) means "take the two smallest (largest)", and *t1, t2* (*t1<t2*) are the time points at which these two lowest (highest) prices are located. If $tg_{t1}^{(n)} < tg_{t2}^{(n)}$ ($pk_{t1}^{(n)} > pk_{t2}^{(n)}$), then the line connecting $tg_{t1}^{(n)}$ and $tg_{t2}^{(n)}$ ($pk_{t1}^{(n)}$ and $pk_{t2}^{(n)}$) is pointing upwards (downwards), so an *uptrend line* is constructed as

$$p_{up}^{(n)}(t) = \left(\frac{tg_{t2}^{(n)} - tg_{t1}^{(n)}}{t2 - t1}\right)t + \left(\frac{tg_{t1}^{(n)} t2 - tg_{t2}^{(n)} t1}{t2 - t1}\right) \quad (23)$$

where $t1 < t2$, $tg_{t1}^{(n)} < tg_{t2}^{(n)}$, and a *downtrend line* is

$$p_{down}^{(n)}(t) = \left(\frac{pk_{t2}^{(n)} - pk_{t1}^{(n)}}{t2 - t1}\right)t + \left(\frac{pk_{t1}^{(n)} t2 - pk_{t2}^{(n)} t1}{t2 - t1}\right) \quad (24)$$

where $t1 < t2, pk_{t1}^{(n)} > pk_{t2}^{(n)}$, time *t* is the horizontal-axis and price $p_t$ is the vertical-axis.

The basic logic behind the trading heuristics related to trend lines is that it is usually difficult to break through a trend line because the existence of a trend line means that a trend has already been in place, so a breakdown (breakup) across the uptrend (downtrend) line will most likely end up with a pullback (throwback). This gives the following heuristic:

**Heuristic 4**: If the current price $p_t$ is getting closer to an uptrend (downtrend) line $p_{up}^{(n)}(t)$ $\left(p_{down}^{(n)}(t)\right)$ from above (below), a buy (sell) order should be placed to bet on the continuation of the uptrend (downtrend).

Define

$$x_{4,t}^{(n)} = \ln\left(p_t / p_{up}^{(n)}(t)\right) \quad (25)$$

$$x_{5,t}^{(n)} = \ln\left(p_t / p_{down}^{(n)}(t)\right) \quad (26)$$

and Heuristic 4 becomes the following *Rule 4 Group*:

Rule $4_1$: IF $x_{4,t}^{(n)}$ is Positive Small (PS), THEN $ed_4$ is Buy Medium (BM)

Rule $4_2$: IF $x_{5,t}^{(n)}$ is Negative Small (NS), THEN $ed_4$ is Sell Medium (SM)  (27)

The excess demand function from the Rule-4-Group traders is the following fuzzy system:

$$ed_4\left(x_{4,t}^{(n)}, x_{5,t}^{(n)}\right) = \frac{0.2\,\mu_{PS}\left(x_{4,t}^{(n)}\right) - 0.2\,\mu_{NS}\left(x_{5,t}^{(n)}\right)}{\mu_{PS}\left(x_{4,t}^{(n)}\right) + \mu_{NS}\left(x_{5,t}^{(n)}\right)} \quad (28)$$

where $0 < x_{4,t}^{(n)} < 2w$ or $-2w < x_{5,t}^{(n)} < 0$, $\mu_{PS}$ and $\mu_{NS}$ are the membership functions of the fuzzy sets PS and NS shown in Fig. 1, and 0.2 and -0.2 are the centers of the fuzzy sets BM and SM shown in Fig. 2.

Fig. 7 shows three simulation runs of the Rule-1+4-Groups case with parameters *(m,n)=(1,5), n\*=100, w=0.01*, $a_1(t) = 0.02$, $a_4(t) = 1$, $\sigma = 0.05$ and $p_0 = 10$, where the first 100 prices came from the pure random walk model (10) and the remaining prices were generated by (1) with all other $a_i(t)$'s except $a_1(t)$ and $a_4(t)$ equal to zero. Comparing Fig. 7 with Fig. 3, we see that more trending happened when the Rule-4-Group traders were added. Originally the Rule-4-Group traders just anticipate that a trend may be coming and act according to their guess, but their actions cause the trend to really happen; this is *self-fulfilling*, a typical phenomenon in real markets.

Another trading heuristic related to trend lines is about trend reversal. Experience has shown that if an established trend line was broken by a relatively large amount, then the trend might have been reversed. This gives the following trading heuristic:

**Heuristic 5:** If the current price $p_t$ is under (above) the uptrend (downtrend) line $y_{up}^{(n)}(t)$ $\left(y_{down}^{(n)}(t)\right)$ for a medium to large amount, then a trend reversal may have been happening so that a small to large sell (buy) order is recommended.

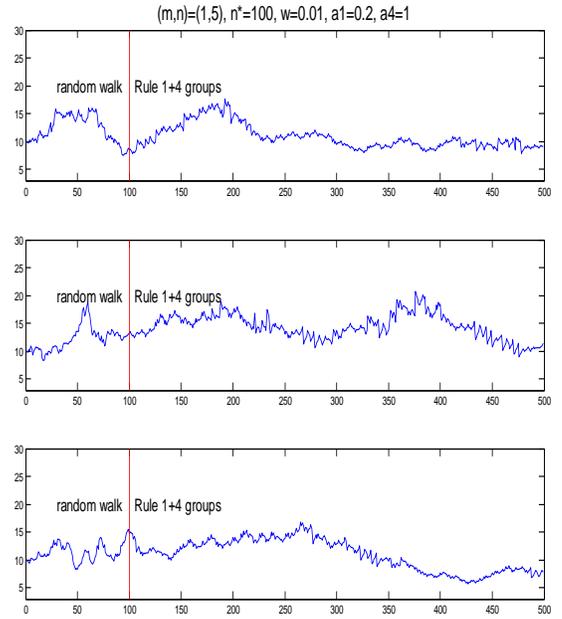

Fig. 7: Three simulation runs of price $p_t$ generated by the random walk model (10) (t=1 to t=100) and the price dynamic model (1) (t=101 to t=500) with the two excess demand functions $ed_1$ and $ed_4$ from Rule 1 and Rule 4 Groups.

This heuristic is transformed into the following *Rule 5 Group*:

Rule $5_1$: IF $x_{4,t}^{(n)}$ is Negative Medium (NM), THEN $ed_5$ is Sell Small (SS)

Rule $5_2$: IF $x_{4,t}^{(n)}$ is Negative Large (NL), THEN $ed_5$ is Sell Big (SB)  (29)

Rule $5_3$: IF $x_{5,t}^{(n)}$ is Positive Medium (PM), THEN $ed_5$ is Buy Small (BS)

Rule $5_4$: IF $x_{5,t}^{(n)}$ is Positive Large (PL), THEN $ed_5$ is Buy Big (BB)

and the excess demand function from the Rule-5-Group traders is the following fuzzy system:

$$ed_5\left(x_{4,t}^{(n)}, x_{5,t}^{(n)}\right) =$$

$$\frac{0.1\,\mu_{PM}\left(x_{5,t}^{(n)}\right) + 0.4\,\mu_{PL}\left(x_{5,t}^{(n)}\right) - 0.1\,\mu_{NM}\left(x_{4,t}^{(n)}\right) - 0.4\,\mu_{NL}\left(x_{4,t}^{(n)}\right)}{\mu_{PM}\left(x_{5,t}^{(n)}\right) + \mu_{PL}\left(x_{5,t}^{(n)}\right) + \mu_{NM}\left(x_{4,t}^{(n)}\right) + \mu_{NL}\left(x_{4,t}^{(n)}\right)}$$

(30)

where $x_{4,t}^{(n)} < -w$ or $x_{5,t}^{(n)} > w$ ($x_{4,t}^{(n)}$ and $x_{5,t}^{(n)}$ are defined in (25) and (26)), PM, PL, NM and NL are fuzzy sets shown in Fig. 1, and 0.1, 0.4, -0.1 and -0.4 are the centers of fuzzy sets BS, BB, SS and SB shown in Fig. 2, respectively.

Fig. 8 shows three simulation runs of the Rule-1+5-Groups case with parameters *(m,n)=(1,5), n\*=100, w=0.01*, $a_1(t) = 0.02$, $a_4(t) = 1$, $\sigma = 0.05$ and $p_0 = 10$, where the first 100 prices came from the pure random walk model (10) and the remaining prices were generated by (1) with all other $a_i(t)$'s except $a_1(t)$ and $a_4(t)$ equal to zero. Again we see self-fulfilling phenomenon in Fig. 8: the big jumps in prices were generated by the Rule-5-group traders who took actions to anticipate the end of the old trends and these actions ended the old trends.

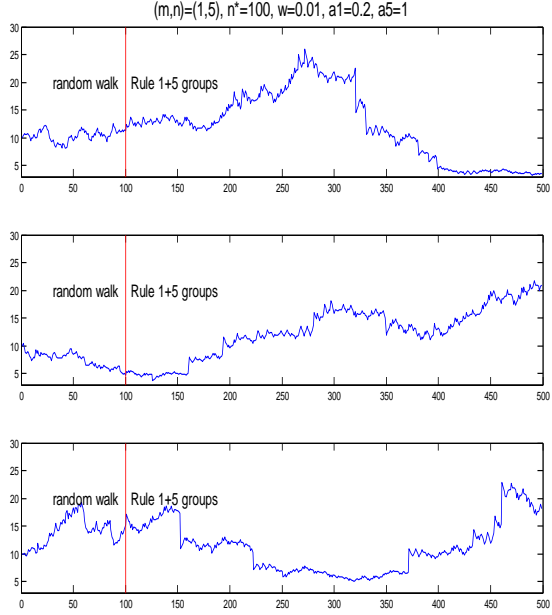

Fig. 8: Three simulation runs for the case of Rule-1-Group plus Rule-5-Group.

## V. BIG BUYER, BIG SELLER AND MANIPULATOR RULES

Big buyers and big sellers are institutional traders who manage large sums of money and usually want to buy or sell a large amount of stocks. Since the amount of stocks offered or asked around the trading price is usually very small, the large buy or sell order has to be cut into small pieces and implemented incrementally over a long period of time (Bouchaud, Farmer and Lillo, 2008; Aldridge, 2013). A reasonable strategy for a big buyer (seller) is to buy (sell) if the price is decreasing (increasing); this gives us the following heuristics: Heuristic 6 for big sellers, and Heuristic 7 for big buyers:

**Heuristic 6:** For a big seller, sell if the price of the stock is increasing: the stronger the increase, the larger the sell order. Keep neutral if the price is decreasing or moving horizontally.

**Heuristic 7:** For a big buyer, buy if the price of the stock is decreasing: the larger the decrease, the larger the buy order. Keep neutral if the price is increasing or moving horizontally.

With $x_{1,t}^{(m,n)}$ defined in (3) and the fuzzy sets in Figs. 1 and 2, we translate Heuristic 6 into the following *Rule 6 Group:*

Rule $6_1$: IF $x_{1,t}^{(1,n)}$ is Positive Small (PS), THEN $ed_6$ is Sell Small (SS)

Rule $6_2$: IF $x_{1,t}^{(1,n)}$ is Positive Medium (PM), THEN $ed_6$ is Sell Medium (SM)

Rule $6_3$: IF $x_{1,t}^{(1,n)}$ is Positive Large (PL), THEN $ed_6$ is Sell Big (SB)  (31)

Rule $6_4$: IF $x_{1,t}^{(1,n)}$ is Around Zero (AZ), THEN $ed_6$ is Neutral (N)

and Heuristic 7 into the following *Rule 7 Group:*

Rule $7_1$: IF $x_{1,t}^{(1,n)}$ is Negative Small (NS), THEN $ed_7$ is Buy Small (BS)

Rule $7_2$: IF $x_{1,t}^{(1,n)}$ is Negative Medium (NM), THEN $ed_7$ is Buy Medium (BM)

Rule $7_3$: IF $x_{1,t}^{(1,n)}$ is Negative Large (NL), HEN $ed_7$ is Buy Big (BB)  (32)

Rule $7_4$: IF $x_{1,t}^{(1,n)}$ is Around Zero (AZ), THEN $ed_7$ is Neutral (N)

The excess demands $ed_6$ and $ed_7$ from Rule 6 Group and Rule 7 Group are

$$ed_6(x_{1,t}^{(1,n)}) = \frac{-0.1\,\mu_{PS}(x_{1,t}^{(1,n)}) - 0.2\,\mu_{PM}(x_{1,t}^{(1,n)}) - 0.4\,\mu_{PL}(x_{1,t}^{(1,n)})}{\mu_{PS}(x_{1,t}^{(1,n)}) + \mu_{PM}(x_{1,t}^{(1,n)}) + \mu_{PL}(x_{1,t}^{(1,n)}) + \mu_{AZ}(x_{1,t}^{(1,n)})} \quad (33)$$

where $x_{1,t}^{(1,n)} > 0$ and

$$ed_7(x_{1,t}^{(1,n)}) = \frac{0.1\,\mu_{NS}(x_{1,t}^{(1,n)}) + 0.2\,\mu_{NM}(x_{1,t}^{(1,n)}) + 0.4\,\mu_{NL}(x_{1,t}^{(1,n)})}{\mu_{NS}(x_{1,t}^{(1,n)}) + \mu_{NM}(x_{1,t}^{(1,n)}) + \mu_{NL}(x_{1,t}^{(1,n)}) + \mu_{AZ}(x_{1,t}^{(1,n)})} \quad (34)$$

where $x_{1,t}^{(1,n)} < 0$, respectively. Figs. 9 and 10 show three simulation runs of the Rule-1+6-Groups and Rule-1+7-Groups cases, respectively, with parameters $(m,n)=(1,5)$, $w=0.01$, $a_1(t) = 0.2$, $a_6(t) = a_7(t) = 0.02$, $\sigma = 0.04$ and $p_0 = 10$, where the first 100 prices came from the pure random walk model (10) and the remaining prices were generated by (1). From Fig. 9 (Fig. 10) we see that the prices are in a general declining (rising) mode when the big sellers (buyers) are presented, although the big sellers (buyers) sell (buy) only when the price is increasing (decreasing).

The study of trading behavior cannot be complete without considering a very important market force: the manipulators. Manipulators are big capitalized traders who use pump-and-dump (Bradshaw, Richardson and Sloan, 2003), spoofing[3], layering[3] (Kirilenko and Lo, 2013) or other strategies (Leinweber and Madhavan, 2001; Fan, 2010; Jiang, 2013) to move the price up or down very quickly to create a mood of optimistic or fear as a trap for other traders. Fig. 11 illustrates a typical working cycle of the pump-and-dump process: in Phase 1 the manipulator trades like a big buyer using Heuristic 7, i.e., collecting a large sum of stocks at relatively low prices; in Phase 2 the manipulator uses strong capital to push the stock price up very quickly (by clearing the ask-side of the order book) to attract the attention of trend-followers; and in Phase 3 the manipulator trades like a big seller using Heuristic 6, i.e., selling the stocks collected in Phase 1 around the new higher prices. Although manipulation is illegal in all stock markets (Kyle and Viswanathan, 2008), it is common practice in almost all markets around the world for hundreds of years (Leinweber and Madhavan, 2001).

---

[3] "Spoofing" involves intentionally manipulating prices by placing an order and then canceling it shortly thereafter, at which point the spoofer consummates a trade in the opposite direction of the canceled order. "Layering" involves placing a successive sequence of limit orders to give the appearance of a change in demand or supply; after a trade is consummated at the manipulated price, the layered limit orders are canceled.

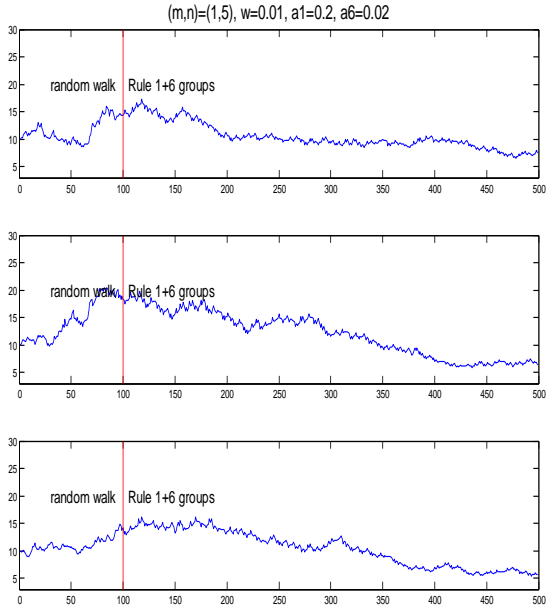

Fig. 9: Three simulation runs for the case of Rule-1-Group plus Rule-6-Group (big sellers).

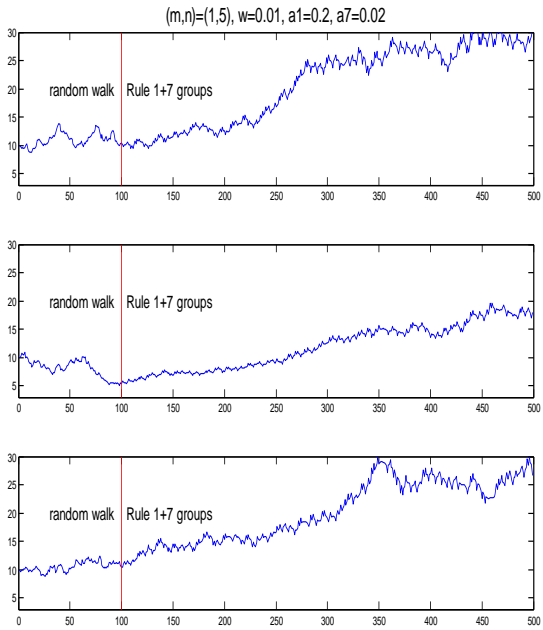

Fig. 10: Three simulation runs for the case of Rule-1-Group plus Rule-7-Group (big buyers).

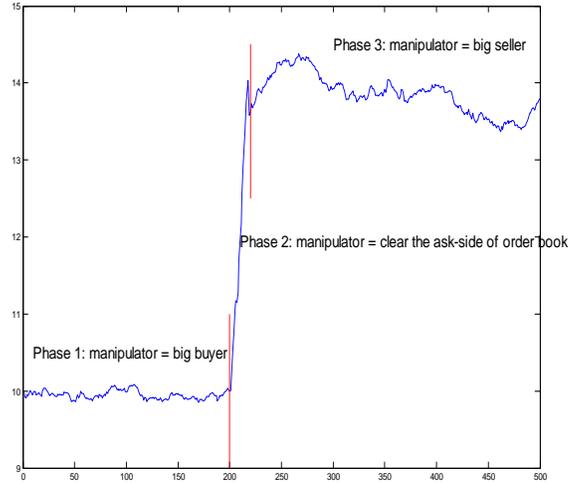

Fig. 11: A typical operation cycle of manipulators: the pump-and-dump process.

Since the manipulator trades like a big buyer (seller) in Phase 1 (Phase 3), the excess demand function equals $ed_7$ ($ed_6$) in these phases. For Phase 2 (the pushing-up phase), the strategy is the following treading heuristic:

**Heuristic 8:** Buy the stock with big money within very short time to clear the ask (sell) side of the order book quickly.

which, when expressed in terms of fuzzy IF-THEN rules, is the following *Rule 8 Group*:

$$\text{Rule 8: Whatever the } x_t\text{'s are, the } ed_8 \text{ is Buy Big (BB)} \quad (35)$$

Putting these three phases together, we obtain the following excess demand function for the manipulator:

$$ed_8\left(x_{1,t}^{(1,n)}\right) =$$

$$\begin{cases} \dfrac{0.1\,\mu_{\text{NS}}\left(x_{1,t}^{(1,n)}\right) + 0.2\,\mu_{\text{NM}}\left(x_{1,t}^{(1,n)}\right) + 0.4\,\mu_{\text{NL}}\left(x_{1,t}^{(1,n)}\right)}{\mu_{\text{NS}}\left(x_{1,t}^{(1,n)}\right) + \mu_{\text{NM}}\left(x_{1,t}^{(1,n)}\right) + \mu_{\text{NL}}\left(x_{1,t}^{(1,n)}\right) + \mu_{\text{AZ}}\left(x_{1,t}^{(1,n)}\right)} & \text{Phase 1} \\ 0.4 & \text{Phase 2} \\ \dfrac{-0.1\,\mu_{\text{PS}}\left(x_{1,t}^{(1,n)}\right) - 0.2\,\mu_{\text{PM}}\left(x_{1,t}^{(1,n)}\right) - 0.4\,\mu_{\text{PL}}\left(x_{1,t}^{(1,n)}\right)}{\mu_{\text{PS}}\left(x_{1,t}^{(1,n)}\right) + \mu_{\text{PM}}\left(x_{1,t}^{(1,n)}\right) + \mu_{\text{PL}}\left(x_{1,t}^{(1,n)}\right) + \mu_{\text{AZ}}\left(x_{1,t}^{(1,n)}\right)} & \text{Phase 3} \end{cases}$$

$$(36)$$

where $x_{1,t}^{(1,n)} < 0$ in Phase 1 and $x_{1,t}^{(1,n)} > 0$ in Phase 3. Fig. 11 is in fact a simulation run of (1) with $ed_1$ and $ed_8$, where Phase 2 is from $t=200$ to $t=220$.

## VI. BAND AND STOP RULES

Bands are envelopes around a moving average with variable size. The basic idea of using bands is that since most prices are contained within the band, a breakout across the boundaries of the band is an indication of the start of a new trend. The most widely used band is the Bollinger Band (Bollinger, 2002) that adds and subtracts the moving estimate of two standard deviations of returns to a moving average. Specifically, let

$$r_t \equiv \ln(p_t) - \ln(p_{t-1}) \cong \frac{p_t - p_{t-1}}{p_{t-1}} \qquad (37)$$

be the return and assume the returns are zero-mean; the moving estimate of the standard deviation of the returns is

$$v_t^{(n)} = \left( \frac{1}{n} \sum_{i=0}^{n-1} r_{t-i}^2 \right)^{1/2} \qquad (38)$$

The upper and lower boundaries of the Bollinger Band are $\bar{p}_{t,n} + 2v_t^{(n)}$ and $\bar{p}_{t,n} - 2v_t^{(n)}$, respectively, where $\bar{p}_{t,n}$ is the price moving average defined in (2). The trading heuristic based on the Bollinger Band is the following:

**Heuristic 9:** When the price $p_t$ is moving out of the Bollinger Band $[\bar{p}_{t,n} - 2v_t^{(n)}, \bar{p}_{t,n} + 2v_t^{(n)}]$, it is a sign of the start of a new trend, so a buy (if the price is above the Band) or sell (if the price is below the Band) order should be in place.

Define

$$x_{6,t}^{(n)} = \ln\left( p_t \big/ (\bar{p}_{t,n} + 2v_t^{(n)}) \right) \qquad (39)$$

$$x_{7,t}^{(n)} = \ln\left( p_t \big/ (\bar{p}_{t,n} - 2v_t^{(n)}) \right) \qquad (40)$$

then Heuristic 9 becomes the following *Rule 9 Group*:

Rule $9_1$: IF $x_{6,t}^{(n)}$ is Positive Small (PS), THEN $ed_9$ is Buy Small (BS)

Rule $9_2$: IF $x_{6,t}^{(n)}$ is Positive Medium (PM), THEN $ed_9$ is Buy Big (BB)

Rule $9_3$: IF $x_{7,t}^{(n)}$ is Negative Small (NS), THEN $ed_9$ is Sell Small (SS) (41)

Rule $9_4$: IF $x_{7,t}^{(n)}$ is Negative Medium (NM), THEN $ed_9$ is Sell Big (SB)

which, when combined into a fuzzy system, give the following excess demand function:

$$ed_9\left(x_{6,t}^{(n)}, x_{7,t}^{(n)}\right) =$$

$$\frac{0.1\,\mu_{PS}\left(x_{6,t}^{(n)}\right) + 0.4\,\mu_{PM}\left(x_{6,t}^{(n)}\right) - 0.1\,\mu_{NS}\left(x_{7,t}^{(n)}\right) - 0.4\,\mu_{NM}\left(x_{7,t}^{(n)}\right)}{\mu_{PS}\left(x_{6,t}^{(n)}\right) + \mu_{PM}\left(x_{6,t}^{(n)}\right) + \mu_{NS}\left(x_{7,t}^{(n)}\right) + \mu_{NM}\left(x_{7,t}^{(n)}\right)} \qquad (42)$$

where $0 < x_{6,t}^{(n)} < 3w$ or $-3w < x_{7,t}^{(n)} < 0$. Fig. 12 shows three simulation runs of the random walk plus Rule-9-Group model:

$$\ln(p_{t+1}) = \ln(p_t) + \sigma\,\varepsilon_t + a_9 ed_9(x_{6,t}^{(n)}, x_{7,t}^{(n)}) \qquad (43)$$

with $n=5$, $w=0.01$, $a_9 = 0.2$, $\sigma = 0.04$ and $p_0 = 10$, where the first 100 prices came from the random walk model (10) and the remaining prices were generated by (43). We see again the self-fulfilling phenomenon: more trending occurred in the model (43) part than in the pure random walk part in Fig. 12, due to the actions for the anticipation of the new trends after the price broke out the band and these actions themselves caused the trends.

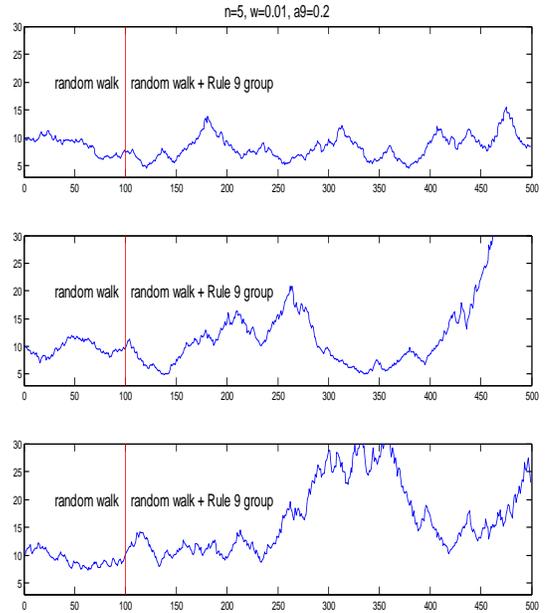

Fig. 12: Three simulation runs of random walk plus Rule-9-Group model (43).

If you asked an investment expert what is the most important element in an investment system, the answer would very likely be: stops. Indeed, making profits is not harmful, whether it is big or small, but you must protect your portfolio from large losses. There are usually two types of stops people use to protect their capital: *protective stops* that set hard limits to the maximum losses allowed and clear the position when the hard limits are reached, and *trailing stops* that protect profits from deteriorating back into a loss. There are many types of protective and trailing stops, such as Hard Money Stops, Breakeven Stops, Technical Point Stops, Volatility Stops, Trend Line Stops, Adaptive Stops, Time Stops, Signal Stops, etc. (Kirkpatrick and Dahlquist, 2011). The following heuristic summarizes a typical protective stop and a typical trailing stop:

**Heuristic 10:** Protective Stop: If a position has a loss approaching 20% and more, it means something was seriously wrong in our previous analysis when we purchased the stock, so the position should be quit as soon as possible. Trailing Stop: If a position has accumulated good profits and the price has dropped around 10% or more from the recent high, then the position should be quit gradually to cash out the profits.

To translate Heuristic 10 into fuzzy IF-THEN rules, let $o_{p_i^{buy}}$ be the amount of a stock in the portfolio purchased at price $p_i^{buy}$ ($i = 1,2,3,...$) and $p_{max}^{(n)}(t) = \max_{t-n+1 \leq k \leq t}\{p_k\}$ be the maximum of the last *n* prices of the stock. Then Heuristic 10 becomes the following *Rule 10 Group*:

Rule 10$_1$: IF $o_{p_i^{buy}}$ is Not Around Zero ($\overline{AZ}$) and $p_t - p_i^{buy}$ is Negative Large (NL), THEN $ed_{10}$ is Sell Big (SB)

Rule 10$_2$: IF $o_{p_i^{buy}}$ is Not Around Zero ($\overline{AZ}$) and $p_t - p_i^{buy}$ is Positive (P) and $p_t - p_{max}^{(n)}(t)$ is Negative Large (NL), THEN $ed_{10}$ is Sell Big (SB)

Rule 10$_3$: IF $o_{p_i^{buy}}$ is Around Zero (AZ), THEN $ed_{10}$ is Neutral (N)  (44)

where P is the union of PS, PM and PL, $\overline{AZ}$ is the complement of AZ, and the parameter *w* in the NL's (see Fig. 1) of Rule 10$_1$ and Rule 10$_2$ are set to 10% and 5%, respectively. Combining the three rules of (44), we obtain the following excess demand function for the stocks purchased at price $p_i^{buy}$:

$$ed_{10i}\left(o_{p_i^{buy}}, p_t, p_i^{buy}, p_{max}^{(n)}(t)\right) =$$

$$\left(-0.4\,\mu_{\overline{AZ}}\left(o_{p_i^{buy}}\right)\mu_{NL}(p_t - p_i^{buy})\right.$$

$$- 0.4\,\mu_{\overline{AZ}}\left(o_{p_i^{buy}}\right)\mu_P(p_t - p_i^{buy})\,\mu_{NL}\left(p_t - p_{max}^{(n)}(t)\right)\right)$$

$$/\left(\mu_{\overline{AZ}}\left(o_{p_i^{buy}}\right)\mu_{NL}(p_t - p_i^{buy})\right.$$

$$+ \mu_{\overline{AZ}}\left(o_{p_i^{buy}}\right)\mu_P(p_t - p_i^{buy})\,\mu_{NL}\left(p_t - p_{max}^{(n)}(t)\right)$$

$$+ \mu_{AZ}\left(o_{p_i^{buy}}\right)\right) \quad (45)$$

If the denominator of (45) equals zero, then $ed_{10i} = 0$. The final excess demand function from Rule-10-Group is the weighted average of the $ed_{10i}$'s with weights $o_{p_i^{buy}}$:

$$ed_{10}\left(o_{p_i^{buy}}, p_t, p_i^{buy}, p_{max}^{(n)}(t)\right) =$$

$$\frac{\sum_i ed_{10i}\left(o_{p_i^{buy}}, p_t, p_i^{buy}, p_{max}^{(n)}(t)\right) o_{p_i^{buy}}}{\sum_i o_{p_i^{buy}}} \quad (46)$$

where $\sum_i o_{p_i^{buy}} \neq 0$ meaning that the amount of this stock in the portfolio is non-zero.

## VII. VOLUME AND STRENGTH RULES

The rules in the last five sections used only the past price information of the same stock. In this section we consider the trading rules that use not only its own past prices but also other information such as volume and the prices of other stocks in the market. Since the key of technical trading is "follow the trend", the indicators considered in this section are used to confirm a trend or to give early warning for a possible trend reversal.

We start with volume. A useful volume indicator is the *on-balance volume* $OBV_t$ defined as follows:

$$OBV_t = V_t\,sign(p_t - p_{t-1}) + OBV_{t-1} \quad (47)$$

where $V_t$ is the traded volume of the stock from *t-1* to *t* and $sign(p_t - p_{t-1})$ is the sign function that equals 1 if $p_t - p_{t-1} > 0$, equals -1 if $p_t - p_{t-1} < 0$ and equals 0 if $p_t - p_{t-1} = 0$. The on-balance volume $OBV_t$ is a measure of the accumulated money inflow into the stock. Therefore it is an early warning of trend reversal if the trend of price $p_t$ and the trend of $OBV_t$ are in opposite directions; this gives the following trading heuristic:

**Heuristic 11:** If the price $p_t$ is in an uptrend (downtrend) but the on-balance volume $OBV_t$ is in a downtrend (uptrend), then as soon as the price shows a small sign of decreasing (increasing), a sell (buy) is recommended to bet on the reversal of the price trend.

Similar to the variables defined in constructing the price trend lines in (21)-(24), we define

$$\left(Vtg_{t1}^{(n)}, Vtg_{t2}^{(n)}\right) =$$

$$\min_{t-n \leq k \leq t-1} 2\{OBV_k | OBV_k < OBV_{k-1}, OBV_k < OBV_{k+1}\} \quad (48)$$

$$\left(Vpk_{t1}^{(n)}, Vpk_{t2}^{(n)}\right) =$$

$$\max2_{t-n \leq k \leq t-1} \{OBV_k | OBV_k > OBV_{k-1}, OBV_k > OBV_{k+1}\} \quad (49)$$

to be the two lowest (highest) troughs (peaks) of the $OBV_t$ in the time interval $[t-n, t-1]$, where $t1, t2$ ($t1<t2$) are the time points at which these two lowest (highest) $OBV_t$'s are located. Let

$$x_{8,t}^{(n)} = \frac{Vtg_{t2}^{(n)} - Vtg_{t1}^{(n)}}{t2 - t1} \quad (50)$$

$$x_{9,t}^{(n)} = \frac{Vpk_{t2}^{(n)} - Vpk_{t1}^{(n)}}{t2 - t1} \quad (51)$$

be the slopes of the lines connecting $Vtg_{t1}^{(n)}$ to $Vtg_{t2}^{(n)}$ and $Vpk_{t1}^{(n)}$ to $Vpk_{t2}^{(n)}$, respectively, we can now transform Heuristic 11 into the following *Rule 11 Group*:

Rule 11$_1$: IF $x_{4,t}^{(n)}$ is NS and $x_{9,t}^{(n)}$ is Negative (N), THEN $ed_{11}$ is SM

Rule 11$_2$: IF $x_{5,t}^{(n)}$ is PS and $x_{8,t}^{(n)}$ is Positive (P), THEN $ed_{11}$ is BM (52)

where the fuzzy set Negative (Positive) is the union of the fuzzy sets NS, NM and NL (PS, PM and PL) with membership function $\mu_N = max\{\mu_{NS}, \mu_{NM}, \mu_{NL}\}$ ( $\mu_P = max\{\mu_{PS}, \mu_{PM}, \mu_{PL}\}$ ). Recalling the definition of $x_{4,t}^{(n)}$ and $x_{5,t}^{(n)}$ in (25) and (26), we see that "$x_{4,t}^{(n)}$ is NS" means "the price has just broken the uptrend line from above" and "$x_{5,t}^{(n)}$ is PS" means "the price has just broken the downtrend line from below". Furthermore, "$x_{9,t}^{(n)}$ is N" ("$x_{8,t}^{(n)}$ is P") confirms the breakdown (breakup) so that Sell Medium (Buy Medium) is executed. The excess demand function from the Rule-11-Group traders is the following fuzzy system:

$$ed_{11}\left(x_{4,5,8,9,t}^{(n)}\right) =$$

$$\frac{0.2 \, \mu_{PS}\left(x_{5,t}^{(n)}\right)\mu_P\left(x_{8,t}^{(n)}\right) - 0.2 \, \mu_{NS}\left(x_{4,t}^{(n)}\right)\mu_N\left(x_{9,t}^{(n)}\right)}{\mu_{PS}\left(x_{5,t}^{(n)}\right)\mu_P\left(x_{8,t}^{(n)}\right) + \mu_{NS}\left(x_{4,t}^{(n)}\right)\mu_N\left(x_{9,t}^{(n)}\right)} \quad (53)$$

If the denominator of (53) equals zero, then $ed_{11} = 0$.

Since volume data are not available for our price models (it is our current research topic to build dynamical models for volumes based on technical trading rules), we will not perform simulations for Rule 11 Group. For real stocks where volume data are available, the Rule-11-Group rules can be incorporated into the price dynamic model (1) in the same manner as the rules in other groups.

Lastly, we consider the trading heuristics concerning how a stock performs comparing to the whole market. Let $index_t$ be some index that measures the overall performance of the market, e.g., it could be the S&P500 for the US stock market or the Shanghai&Shenzhen 300 for China stock market. The *relative strength $RS_t$* of a stock with price $p_t$ is defined as

$$RS_t = {p_t}/{index_t} \quad (54)$$

If $RS_t$ is rising (declining), it means the stock outperforms (underperforms) the market average. Since the market has the tendency of stronger getting stronger and weaker getting weaker (Jegadeesh and Titman, 1993), a breakup (breakdown) of the downtrend (uptrend) line, if accompanied by a rising (declining) $RS_t$, is a strong sign of trend reversal. This gives us the following trading heuristic:

**Heuristic 12:** If the price $p_t$ is in an uptrend (downtrend) but its relative strength $RS_t$ is in a downtrend (uptrend), then as soon as the price shows a small sign of decreasing (increasing), a sell (buy) is recommended to leave (enter) this weak (strong) stock.

Define the log-ratio of $RS_t$ to its *n*-leg moving average as

$$x_{10,t}^{(n)} = \ln\left(RS_t \Big/ \left(\frac{1}{n}\sum_{i=0}^{n-1} RS_{t-i}\right)\right) \quad (55)$$

Heuristic 12 gives the following *Rule 12 Group*:

Rule 12$_1$: IF $x_{4,t}^{(n)}$ is NS and $x_{10,t}^{(n)}$ is Negative (N), THEN $ed_{12}$ is SM

Rule 12$_2$: IF $x_{5,t}^{(n)}$ is PS and $x_{10,t}^{(n)}$ is Positive (P), THEN $ed_{12}$ is BM (56)

and its excess demand function is

$$ed_{12}\left(x_{4,t}^{(n)}, x_{5,t}^{(n)}, x_{10,t}^{(n)}\right) =$$

$$\frac{0.2 \, \mu_{PS}\left(x_{5,t}^{(n)}\right)\mu_P\left(x_{10,t}^{(n)}\right) - 0.2 \, \mu_{NS}\left(x_{4,t}^{(n)}\right)\mu_N\left(x_{10,t}^{(n)}\right)}{\mu_{PS}\left(x_{5,t}^{(n)}\right)\mu_P\left(x_{10,t}^{(n)}\right) + \mu_{NS}\left(x_{4,t}^{(n)}\right)\mu_N\left(x_{10,t}^{(n)}\right)} \quad (57)$$

where $ed_{12} = 0$ if the denominator of (57) equals zero. The price dynamics driven by $ed_{12}$ can be simulated if we consider a multi-stock simulation model; however, we will leave it to another paper concentrating on this very important topic: interactions among multiple stocks.

## VIII. CONCLUDING REMARKS

Stock prices are generated by the actions of stock traders, and according to a survey of 692 fund managers in five countries

(Menkhoff, 2010), the vast majority of the fund managers rely on technical analysis[4]. The main contribution of this paper is to build a bridge between the fuzzy, linguistic world of technical analysis and the rigorous, mathematical world of nonlinear dynamic equations, as illustrated in Fig. 13. In Part I of this paper, we developed the details of transforming twelve common trading heuristics into nonlinear dynamical equations and simulated these models to illustrate how simple technical trading rules can produce complex price patterns. There have been a huge number of technical trading rules accumulated during the long history of technical analysis (Lo and Hasanhodzic, 2010) and more rules are being proposed constantly to adapt to the evolving market conditions[5]; by using the methods of this paper we can transform these linguistic descriptions of trading wisdoms into dynamical equations and observe exactly what price trajectories and patterns these heuristic rules can produce. Furthermore, these dynamical equations allowed us to study the price dynamics generated by the technical trading rules in a mathematically rigorous fashion (as we will do in Part II of this paper) and push technical analysis forward in the direction from an art to a science. Practically, these models provide us a framework to detect the hidden operations in the market so that trading strategies can be developed to beat the benchmark Buy-and-Hold strategy (as we will do in Part III of this paper).

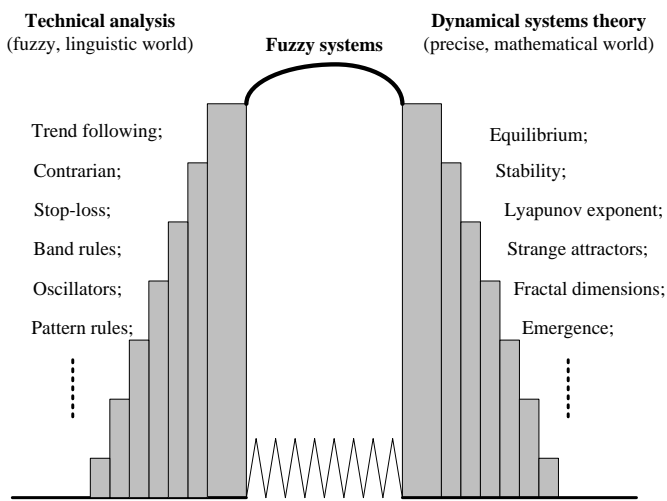

Fig. 13: The "main story" this paper is trying to tell: A bridge is being built across the gulf between the fuzzy, linguistic world of technical analysis which is rich in trading wisdom on one side and the precise, mathematical world of dynamical systems theory which is rich in rigorous tools on the other side.

---

[4] More specifically, for the forecasting horizon of weeks, the relative importance is 29.4 for Technical Analysis, 28.4 for Order Flow, and 1.4 for Fundamental Analysis for US fund managers; for German fund managers, these numbers are 60 for Technical Analysis, 22.6 for Order Flow, and 6.7 for Fundamental Analysis; see Fig. 2 in Menkhoff 2010 for more data.

[5] According to Schmidt 2011, a search on Amazon.com using the key words *technical analysis* for new books that were printed between January 2001 and September 2010 yielded 468 entries --- roughly one new book every week.

The price dynamical models proposed in this paper are totally deterministic, so a natural question is how these models compare with the mainstream stochastic models that provide measures of unpredictable uncertainties[6]? To answer this question, let's see how the mainstream stochastic models handle the unpredictable uncertainties: these models usually have a simple deterministic part (linear or simple ad-hoc nonlinear functions of state variables), and leave the real important part of the (unknown) dynamics to the noise term (representing unpredictable uncertainties); then, they assume that the noise term is zero-mean and i.i.d. (or make other comparable assumptions in order for the mathematics to march forward beautifully). We know a key characteristic of financial systems is evolving (time-varying, non-stationary[7]; Lo, 2004). How can one faithfully assume a key part of an evolving, non-stationary dynamics (the unpredictable uncertainties) to be zero-mean (a time-invariant constant)? Furthermore, the i.i.d. assumption implies that the traders are drunkards (Malkiel, 2012) --- the move in one step has no relation to the moves of the previous steps. Real traders are betting in real money, they are serious, they are watching the evolving markets closely and making decisions accordingly, they are socially connected and influence each other, and they are not some random noise from the background as in an electronic circuit model. So, we will ultimately have to roll up our sleeves to dig deep into the origin of the price formation machine[8] to build a deterministic, cause-effect model for the key dynamics; the models proposed in this paper are a trial in this direction.

The main difference between Natural Sciences and Social Sciences is that we know the basic laws governing the motions of the objects in Natural Sciences such as the Newton's Laws, whereas the basic objects in Social Sciences are human beings and we do not have the "Newton's Laws for human subjects." That is, we lack a deep underpinning for understanding the mechanisms at the origin of the dynamical behavior of financial markets (Mirowski, 1989; Kirman, 1989; Malevergne and Sornette, 2006). Our current stage of understanding social systems is quite similar to "the blind monks touching the elephant". In our pursuit for the basic laws of stock markets, we should listen carefully to what the real traders say about their crafts (e.g. Livermore, 1940; Lynch, 1989; Soros, 2003; Lindsey and Schachter, 2007; Lo and Hasanhodzic, 2009), because human traders are the "atoms of stock markets." As one of the most successful speculators in modern finance, George Soros has a Theory of Reflexivity which, in his own words (page 2 of Soros, 2003), is described as follows:

> The concept of reflexivity is very simple. In situations that have thinking participants, there is a two-way interaction between the participant's thinking and the situation in which they participate. On the one hand, participants seek to understand reality; on the other hand, they seek to bring about a desired outcome. The two functions work in opposite directions: in the cognitive function reality is the given; in the participating function, the participants' understanding is the constant. The two functions can interfere with each other by rendering what is supposed to be given, contingent. I call the interference between the two functions "reflexivity."

---

[6] "The human brain is wonderful at spotting patterns. It's an ability that is one of the foundation stones of science. When we notice a pattern, we try to pin it down mathematically, and then use the maths to help us understand the world around us. And if we can't spot a pattern, we don't put its absence down to ignorance. Instead we fall back on our favourite alternative. We call it randomness." --- Ian Stewart (Stewart, 2004).

[7] Traders come and go (Foucault, Pagano and Roell, 2013), and different traders use different trading strategies, which results in different price patterns at different time points, meaning that the returns are non-stationary.

[8] By which we mean both the mechanical setup of the limit order book dynamics (Gould, Porter, Williams, McDonald, Fenn and Howison, 2013; Bouchaud, Farmer and Lillo, 2008; Cont, Kukannov and Stoikov, 2014) and the investors' mindset (Keynes, 1935; Graham, 1973; Lynch, 1989; Soros, 2003; etc.) including their "obstinate passion" for technical analysis (Menkhoff and Taylor, 2007).

To represent our price dynamical model (1) in the framework of Soros's Reflexivity Theory, we plot the block diagram of model (1) in Fig. 14, where we view the excess demand $ed_i$ of each trader group as the cognitive function in Soros's reflexivity theory and the time-varying relative strength $a_i(t)$ as the participating function. According to the concept of reflexivity cited above, the input to the cognitive function $ed_i$ is the reality --- the price history and other information, and the output is $ed_i$ which can be viewed as the dollar amount to trade the stock: a positive $ed_i$ means to buy the stock with $ed_i$ dollars, and a negative $ed_i$ means to sell the stock to cash out $|ed_i|$ dollars. The impact of the dollar amount $ed_i$ on the price dynamics is through the participating function $a_i(t)$ whose exact value cannot be specified by the Rule-$i$-Group traders alone, because the impact depends on the actions of other traders that are beyond the control of the Rule-$i$-Group traders. In this framework, reflexivity means the interference between $a_i(t)$ and the internal mechanism of $ed_i$ --- the fuzzy trading rules and their parameters $m$, $n$ and $w$; this gives us some hints for possible directions of future research.

Indeed, in Part III of this paper we will develop two trading strategies --- Follow-the-Big-Buyer and Ride-the-Mood --- based on the estimation of the $a_i(t)$'s and test the strategies over the top 20 blue-chip stocks in the Hong Kong Stock Exchange to show the superior performance of these trading strategies over the benchmark Buy-and-Hold strategy (Part III can be found at arXiv:1401.1892). But before we do this, we would better first analyze the price dynamical model (1) in detail to get a basic understanding for the key properties of the price dynamics, which is what we are going to do in Part II of this paper (Part II can be found at arXiv:1401.1891), next.

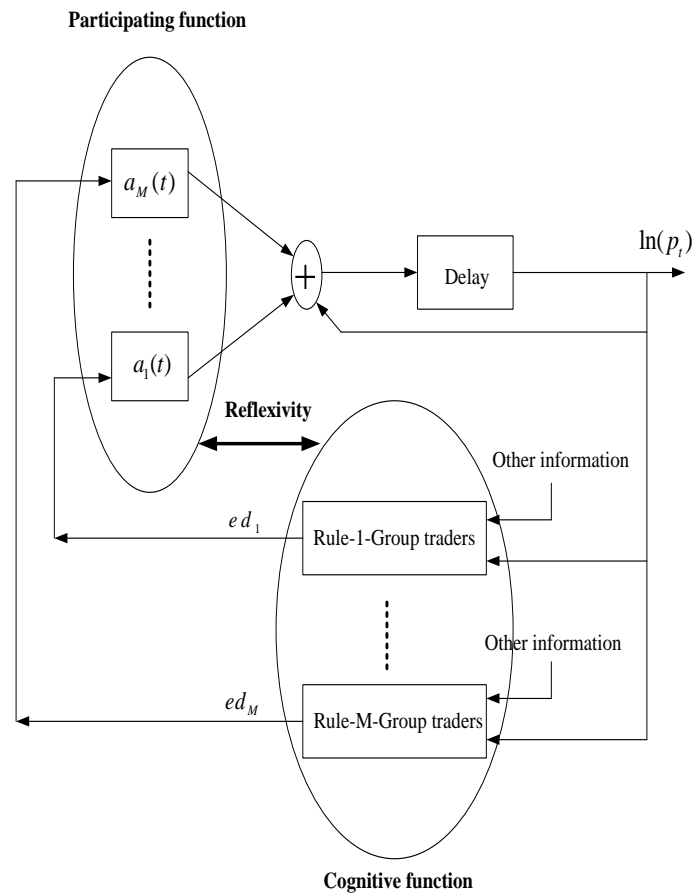

Fig. 14: The block diagram representation of the price dynamical model (1) with interpretation of the function blocks in terms of George Soros's Theory of Reflexivity.

ACKNOWLEDGMENT

The author would like thank the anonymous reviewers for their very insightful and inspiring comments that helped to improve the paper.

REFERENCES

[1] Aldridge, I., *High-Frequency Trading: A Practical Guide to Algorithmic Strategies and Trading Systems (2nd Edition),* John Wiley & Sons, Inc., New Jersey, 2013.
[2] Andersen, T.G., T. Bollerslev and F.X. Diebold, "Parametric and Nonparametric Volatility Measurement," in Y. Ait-Sahalia and L.P. Hansen (ed.), *Handbook of Financial Econometrics: Tools and Techniques*, Vol. 1, pp. 67-138, Elsevier: North-Holland, 2010.
[3] Atsalakis, G. and K. Valavanis, "Surveying Stock Market Forecasting Techniques - Part II: Soft Computing Methods," *Expert Systems with Applications* 36: 5932-5941, 2008.
[4] Bachelier, L., "Theorie de la speculation." Thesis. Annales Scientifiques de l'Ecole Normale Superieure, 1990. III-17, 21-86. Translated by Cootner (ed.) as "Random character of Stock Market Prices." (Massachusetts Institute of Technology, 1964: 17-78).
[5] Black, F. and M. Scholes, "The Pricing of Options and Corporate Liabilities," *Journal of Political Economy* 81: 637-659, 1973.

[6] Bollerslev, T., R.Y. Chou and K.F. Kroner, "ARCH Modeling in Finance: A Review of the Theory and Empirical Evidence," *Journal of Econometrics* 52: 5-59, 1992.

[7] Bollinger, J., *Bollinger on Bollinger Bands*, McGraw-Hill, New York, 2002.

[8] Bouchaud, J.P., D. Farmer and F. Lillo, "How Markets Slowly Digest Changes in Supply and Demand," in T. Hens, K. Schenk-Hoppe, eds., *Handbook of Financial Market: Dynamics and Evolution*, Elsevier: Academic Press: *57-160*, 2008.

[9] Bouchaud, J.P. and M. Potters, *Theory of Financial Risks: From Statistical Physics to Risk Management (2$^{nd}$ Edition)*, Cambridge University Press, Cambridge, 2003.

[10] Bradshaw, M.T., S.A. Richardson and R.G. Sloan, "Pump and Dump: An Empirical Analysis of the Relation Between Corporate Financing Activities and Sell-side Analyst Research," *SSNR 410521*, 2003.

[11] Brock, W.A. J. Lakonishok and B. LeBaron, "Simple Technical Trading Rules and the Stochastic Properties of Stock Returns," *Journal of Finance* 47: 1731-1764, 1992.

[12] Cont, R., A. Kukannov and S. Stoikov, "The Price Impact of Order Book Events," *J. Financial Econometrics* 12: 47-88, 2014.

[13] Elton, E.J., M.J. Gruber, S.J. Brown and W.N. Goetzmann, *Modern Portfolio Theory and Investment Analysis (7$^{th}$ Edition)*, John Wiley & Sons, Inc., 2007.

[14] Engle, R.F., "Autoregressive Conditional Heteroskedasticity with Estimates of the Variance of U.K. Inflation," *Econometrica* 50: 987-1008, 1982.

[15] Fama, E., "Efficient Capital Markets: A Review of Theory and Empirical Work," *Journal of Finance* 25: 383-417, 1970.

[16] Fan, J.J., *Follow the Banker (in Chinese)*, China Machine Press, Beijing, 2010.

[17] Foucault, T., M. Pagano and A. Roell, *Market Liquidity: Theory, Evidence, and Policy*, Oxford University Press, 2013.

[18] Fouque, J.P., G. Papanicolaou and K.R. Sircar, *Derivatives in Financial Markets with Stochastic Volatility*, Cambridge University Press, 2000.

[19] Gencay, R., "The Predictability of Security Returns with Simple Technical Trading Rules," *Journal of Empirical Finances* 5: 347-359, 1998.

[20] Gigerenzer, G., "On Narrow Norms and Vague Heuristics: A Reply to Kahneman and Tversky (1996)," *Psychological Review* 103: 592-596, 1996.

[21] Gigerenzer, G. and W. Gaissmaier, "Heuristic Decision Making," *Annual Review of Psychology* 62: 451-482, 2011.

[22] Gould, M.D., M.A. Porter, S. Williams, M. McDonald, D.J. Fenn and S.D. Howison, "Limit Order Books," *Quantitative Finance* 13: 1709-1742, 2013.

[23] Graham, B., *The Intelligent Investor (Revised Edition),* Harper Business, 1973.

[24] Graham, B. and D. Dodd, *Security Analysis (6$^{th}$ Edition),* (McGraw-Hill, New York, 2009), 1940.

[25] Hommes, C.H., "Heterogeneous Agent Models in Economics and Finance," in L. Tesfatsion and K.L. Judd, eds., *Handbook of Computational Economics Vol. 2: Agent-Based Computational Economics*, Elsevier B.V.: 1109-1186, 2006.

[26] Hull, J.C., *Options, Futures, and Other Derivatives (7$^{th}$ Edition)*, Pearson Education, Inc., 2009.

[27] Jegadeesh, N and S. Titman, "Returns to Buying Winners and Selling Losers: Implications for Stock Market Efficiency," *Journal of Finance* 48: 65-91, 1993.

[28] Jiang, X.S., *Reveal the Secrets of Main Force Manipulation (in Chinese)*, Tsinghua University Press, 2013.

[29] Keynes, J.M., *The General Theory of Employment, Interest and Money*, (BN Publishing, 2009), 1935.

[30] Kirman, A., "The Intrinsic Limits of Modern Economic Theory: The Emperor Has No Clothes," *Economic Journal* 99:126-139, 1989.

[31] Kirilenko, A.A. and A.W. Lo, "Moore's Law versus Murphy's Law: Algorithmic Trading and Its Discontents," *Journal of Economic Perspectives*, 27: 51-72, 2013.

[32] Kirkpatrick, C.D. and J. Dahlquist, *Technical Analysis: The Complete Resource for Financial market Technicians (Second Edition),* Pearson Education, New Jersey, 2011.

[33] Kirkpatrick, C.D. and J. Dahlquist, *Study Guide for the Second Edition of Technical Analysis: The Complete Resource for Financial market Technicians,* Pearson Education, New Jersey, 2013.

[34] Kyle, A.S., "Continuous Auction and Insider Trading," *Econometrica* 53: 1315-1335, 1985.

[35] Kyle, A.S. and S. Viswanathan, "Price Manipulation in Financial Markets: How to Define Illegal Price Manipulation," *American Economic Review* 98:2, 274-279, 2008.

[36] LeBaron, B., S.H. Chen and S. Sunder, "The Future of Agent-Based Research in Economics: A Panel Discussion," *Eastern Economic Journal* 34: 550-565, 2008.

[37] Leinweber, D.J. and A.N. Madhavan, "Three Hundred Years of Stock Market Manipulations," *The Journal of Investing* 10: 7-16, 2001.

[38] Lindsey, R.R. and B. Schachter, *How I Became a Quant: Insights From 25 of Wall Street's Elite*, John Wiley & Sons, New Jersey, 2007.

[39] Livermore, J., *How to Trade in Stocks*, (McGraw-Hill, 2001), 1940.

[40] Lo, A.W., "The Adaptive Market Hypothesis: Market Efficiency from an Evolutionary Perspective," *The Journal of Portfolio Management* 30: 15-29, 2004.

[41] Lo, A.W. and J. Hasanhodzic, *The Heretics of Finance: Conversations with Leading Practitioners of Technical Analysis*, Bloomberg Press, New York, 2009.

[42] Lo, A.W. and J. Hasanhodzic, *The Evolution of Technical Analysis: Financial Prediction from Babylonian Tablets to Bloomberg Terminals*, John Wiley & Sons, Inc., New Jersey, 2010.

[43] Lo, A.W., H. Mamaysky and J. Wang, "Foundations of Technical Analysis: Computational Algorithms, Statistical Inference, and Empirical Implementation," *Journal of Finance* 55: 1705-1765, 2000.

[44] Lux, T., "The Socio-Economic Dynamics of Speculative Markets: Interacting Agents, chaos, and the Fat Tails of Return Distribution," *Journal of Economic Behavior & Organizations* 33: 143-165, 1998.

[45] Lux, T. and M. Marchesi, "Scaling and Criticality in A Stochastic Multi-Agent Model of A Financial Market," *Nature* 397: 498-500, 1999.

[46] Lynch, P., *One Up on Wall Street: How to Use What You Already Know to Make Money in the Market*, Simon & Schuster Paperbacks, New York, 1989.

[47] Malevergne, Y. and D. Sornette, *Extreme Financial Risks: From Dependence to Risk Management*, Springer-Verlag Berlin Heidelberg, 2006.

[48] Malkiel, B.G., *A Random Walk Down Wall Street (10$^{th}$ Edition)*, W.W. Norton & Co., New York, 2012.

[49] Menkhoff, L., "The Use of Technical Analysis by Fund Managers: International Evidence," *Journal of Banking & Finance* 34: 2573-2586, 2010.

[50] Menkhoff, L. and M.P. Taylor, "The Obstinate Passion of Foreign Exchange Professionals: Technical Analysis," *Journal of Economic Literature* XLV: 936-972, 2007.

[51] Merton, R.C., "Option Pricing When Underlying Stock Returns Are Discontinuous," *Journal of Financial Economics* 3: 125-144, 1976.

[52] Mirowski, P., *More Heat Than Light: Economics as Social Physics, Physics as Nature's Economics*, Cambridge University Press, 1989.

[53] Schmidt, A.B., *Financial Markets and Trading: An Introduction to Market Microstructure and Trading Strategies*, John Wiley & Sons, New Jersey, 2011.

[54] Soros, G., *The Alchemy of Finance*, John Wiley & Sons, New Jersey, 2003.

[55] Stewart, I., "In the Lap of the Gods," *NewScientist*, 25 Sept. 2004.

[56] Sullivan, R., A. Timmermann and H. White, "Data-Snooping, Technical Trading Rule Performance, and the Bootstrap," *The Journal of Finance* 54: 1647-1691, 1999.
[57] Tesfatsion, T. and K.L. Judd, eds., *Handbook of Computational Economics Vol. 2: Agent-Based Computational Economics*, Elsevier B.V., 2006.
[58] Wang, L.X., *A Course in Fuzzy Systems and Control*, Prentice-Hall, New Jersey, 1997.
[59] Zadeh, L.A., "Toward a Theory of Fuzzy Systems," in *Aspects of Network and System Theory*, eds. R.E. Kalman and N. DeClaris, Rinehart and Winston, New York, 1971.
[60] Zadeh, L.A., "Outline of a New Approach to the Analysis of Complex Systems and Decision Processes," *IEEE Trans. on Systems, Man, and Cybern.* 3: 28-44, 1973.

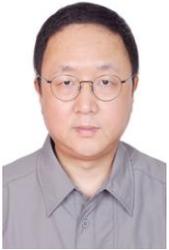

**Li-Xin Wang** received the Ph.D. degree in 1992 from the Department of Electrical Engineering – Systems, University of Southern California (won USC's Phi Kappa Phi's highest Student Recognition Award). From 1992 to 1993 he was a Postdoc Fellow with the Department of Electrical Engineering and Computer Science, University of California at Berkeley. From 1993 to 2007 he was on the faculty of the Department of Electronic and Computer Engineering, The Hong Kong University of Science and Technology. In 2007 he resigned from his tenured position at HKUST to become an independent researcher and investor in the stock and real estate markets in Hong Kong and China. He returned to academia in Fall 2013 by joining the faculty of the Department of Automation Science and Technology, Xian Jiaotong University, after a fruitful hunting journey across the wild land of investment to achieve financial freedom. His research interests are dynamical models of asset prices, market microstructure, trading strategies, fuzzy systems, and opinion networks.